\documentclass[11pt,a4paper,twoside]{article}
\pdfoutput=1

\usepackage[a4paper,left=0.9in,right=0.9in,top=0.85in,bottom=1in,marginparwidth=0pt]{geometry}
\usepackage{newpxtext}
\usepackage[vvarbb]{newpxmath}
\usepackage{amsmath,amsthm,mathtools}
\usepackage{thm-restate}
\usepackage{xspace}
\usepackage{tikz}
\usetikzlibrary{arrows.meta,decorations.pathreplacing,positioning}

\usepackage[T1]{fontenc}
\usepackage[utf8]{inputenc}

\usepackage[tracking=true,protrusion=true,expansion=true]{microtype}
\usepackage{enumitem}
\setlist{itemsep=2pt,topsep=5pt,parsep=0pt}
\usepackage{comment}

\usepackage{xcolor}
\definecolor{accent}{HTML}{1D4E89}
\definecolor{accentmuted}{HTML}{5B84B1}
\definecolor{ink}{HTML}{1A1A1A}
\definecolor{muted}{HTML}{5A5A5A}
\color{ink}

\newtheoremstyle{modernplain}{0.9\baselineskip}{0.9\baselineskip}{\itshape}{0pt}{\bfseries\color{accent}}{.}{0.6em}{}
\newtheoremstyle{moderndef}{0.9\baselineskip}{0.9\baselineskip}{\normalfont}{0pt}{\bfseries\color{accent}}{.}{0.6em}{}
\theoremstyle{modernplain}
\usepackage{thm-restate}
\usepackage{thmtools}

\usepackage{titlesec}
\titleformat{\section}{\normalfont\Large\bfseries\color{accent}}{\thesection}{0.6em}{}
\titleformat{\subsection}{\normalfont\large\bfseries\color{accent}}{\thesubsection}{0.6em}{}
\titleformat{\paragraph}[runin]{\normalfont\bfseries\color{accent}}{}{0pt}{}[\ ]
\titlespacing*{\section}{0pt}{1.8em}{0.8em}
\titlespacing*{\subsection}{0pt}{1.4em}{0.6em}
\titlespacing*{\paragraph}{0pt}{1.4em}{0.8em}

\usepackage[labelfont={bf,color=accent},font=small,labelsep=period]{caption}

\usepackage{fancyhdr}
\fancypagestyle{plain}{%
  \fancyhf{}
  \fancyfoot[C]{\footnotesize\color{muted}\thepage}
  
}

\makeatletter
\renewenvironment{proof}[1][\proofname]{\par\pushQED{\qed}%
  \normalfont\topsep6\p@\@plus6\p@\relax
  \trivlist
  \item[\hskip\labelsep\color{accent}\bfseries #1\@addpunct{.}]\ignorespaces
}{\popQED\endtrivlist}
\makeatother

\usepackage{xspace}
\usepackage{xparse}
\usepackage{algorithm}
\usepackage{mathtools}
\usepackage[commentColor=black]{algpseudocodex}

\makeatletter
\newcommand{\@authorlist}{}
\newcounter{authorcount}
\RenewDocumentCommand{\author}{m g g}{%
  \ifnum\value{authorcount}>0 \g@addto@macro\@authorlist{\hspace{2.5em}}\fi
  \stepcounter{authorcount}%
  \g@addto@macro\@authorlist{%
    \begin{minipage}[t]{0.42\textwidth}\centering
      \textbf{#1}%
      \IfNoValueTF{#2}{}{\\[3pt]{\small\itshape #2}}%
      \IfNoValueTF{#3}{}{\\[2pt]{\small\href{mailto:#3}{\nolinkurl{#3}}}}%
    \end{minipage}%
  }%
}
\newcommand{\@fundingextra}{}
\newcommand{\keywords}[1]{\renewcommand{\@keywordstext}{#1}}
\newcommand{\@keywordstext}{}
\newcommand{\funding}[1]{\renewcommand{\@fundingtext}{#1}}
\newcommand{\@fundingtext}{}

\renewcommand{\maketitle}{%
  \begingroup
  \thispagestyle{plain}
  \begin{center}
    {\fontsize{19}{23}\selectfont\bfseries\color{accent}\@title\par}
    \vspace{1.3em}
    {\setlength{\lineskip}{1.6em}\@authorlist\par}
  \end{center}
  \vspace{1.3em}
  {\color{accent}\hrule height0.6pt}
  \endgroup
}

\renewenvironment{abstract}{%
  \vspace{1.2em}
  \noindent{\bfseries\scshape\color{accent}\small Abstract}\par\vspace{0.5em}
  \small\noindent\ignorespaces
}{%
  \par\vspace{1.2em}
  \small
  \ifx\@keywordstext\@empty\else\noindent{\bfseries\color{accent}Keywords.}~\@keywordstext\par\vspace{4pt}\fi
  \ifx\@fundingtext\@empty\else\noindent{\bfseries\color{accent}Funding.}~\@fundingtext\@fundingextra\par\vspace{4pt}\fi
  \vspace{0.6em}
  {\color{accent}\hrule height0.6pt}
  \vspace{1.6em}
}
\makeatother

\usepackage[colorlinks,linkcolor=accent,citecolor=accent,urlcolor=accent,anchorcolor=accent]{hyperref}

\colorlet{figA}{accent}
\definecolor{figB}{HTML}{C8551A}

\usepackage{hyperref}
\usepackage[nameinlink,capitalise]{cleveref}
\crefname{claim}{Claim}{Claims}

\newtheorem{theorem}{Theorem}
\newtheorem{lemma}[theorem]{Lemma}
\newtheorem{corollary}[theorem]{Corollary}

\newtheorem{definition}[theorem]{Definition}

\newcommand{\define}{:=}
\newcommand{\size}[1]{\lVert #1 \rVert}
\newcommand{\vm}{\mathbf{m}}
\newcommand{\Opt}{\text{\textsc{Opt}}\xspace}
\newcommand{\HoH}{\text{\textsc{Hoh}}\xspace}
\newcommand{\vy}{\mathbf{y}}
\newcommand{\va}{\mathbf{a}}
\newcommand{\zero}{\mathbf{0}}
\newcommand{\gs}{g^\star}
\newcommand{\deff}{d}
\newcommand{\real}{\mathbb{R}}
\newcommand{\suchthat}{\;:\;}
\DeclarePairedDelimiter{\set}{\{}{\}}

\title{A 1.764-competitive algorithm for online fractional vertex cover on forests}
\author{Marcin Bienkowski\textsuperscript{$*$}}{University of Wroc{\l}aw, Poland}{marcin.bienkowski@cs.uni.wroc.pl}
\date{\today}

\begin{document}
\maketitle \begingroup\renewcommand{\thefootnote}{$*$}\footnotetext{This work has been supported by Polish National Science Centre grant 2022/45/B/ST6/00559.}\endgroup

\begin{abstract}
We study online fractional vertex cover on forests under edge arrival. We give a deterministic algorithm with competitive ratio at most $4 - \sqrt{5} \approx 1.764$, which simultaneously produces an online fractional matching with competitive ratio of at least $1/(4 - \sqrt{5}) \approx 0.566$. This improves on the previous best competitive ratios of $11/6 \approx 1.833$ and $5/9 \approx 0.555$, respectively.
\end{abstract}


\section{Introduction}
\label{sec:intro}

In the \emph{online fractional vertex cover problem with edge arrivals}, a graph is revealed one edge at a time, for a fixed vertex set~$V$. An algorithm maintains a non-decreasing vector of weights $\vy : V \to \real_{\geq 0}$, initially $\vy = \zero$. For each revealed edge $(u,v)$, the algorithm has to ensure that $y_u + y_v \geq 1$; to this end, the algorithm may increase arbitrary weights. The goal is to minimize the sum of all weights.

We focus on the \emph{online} model, where the algorithm sees only the edges revealed so far. For a~vector~$\vy$, we write $\size{\vy} \define \sum_x y_x$ for its \emph{weight}. We write $\vy^t$ for the vector maintained after step~$t$ and $\Opt_t$ for the minimum weight of a~fractional vertex cover of the edges revealed so far. The competitive ratio of an algorithm is the supremum of $\max_{t \geq 1} \size{\vy^t}/\Opt_t$ over instances.

\paragraph{Online fractional matching.}

Similarly, in the \emph{online fractional matching problem with edge arrivals}, an~algorithm maintains a~non-decreasing vector of weights $\vm : E \to \real_{\geq 0}$, initially $\vm = \zero$. For each revealed edge $(u,v)$, the~algorithm may increase $m_{(u,v)}$ arbitrarily, but it must ensure that for every vertex~$x$, the sum of weights of edges incident to~$x$ is at most~$1$. The goal is to maximize the sum of all weights. Analogously, the competitive ratio of an algorithm is the infimum of $\min_{t \geq 1} \size{\vm^t}/\Opt_t$ over instances. By the strong LP duality, the values of optimal solutions $\Opt_t$ for the two problems are equal.

\paragraph{Previous work.}

For general graphs, the online complexity of fractional vertex cover and fractional matching is resolved. The trivial algorithm that takes both endpoints of every revealed edge is $2$-competitive for vertex cover, and the greedy algorithm is $1/2$-competitive for matching. Tang and Zhang~\cite{TanZha26} showed that no algorithm for vertex cover is better than $2$-competitive, and Gamlath, Kapralov, Maggiori, Svensson and Wajc~\cite{GaKMSW19} showed that no algorithm for matching is better than $1/2$-competitive.

This motivated the study of the problem on forests. For the fractional vertex cover, Balig{\'a}cs, Bosek, Disser, Feldmann, Gutowski, K{\k{e}}pi{\'n}ska, Putra and Zych-Pawlewicz~\cite{BBDFGK26} gave an~algorithm RGB with competitive ratio $11/6 \approx 1.833$, and showed a lower bound of $1.5$ holding even in the so-called \emph{incremental} model. For the fractional matching, Buchbinder, Segev and Tkach~\cite{BuSeTk19} gave an algorithm with ratio of $5/9 \approx 0.555$, and showed an upper bound of $0.591$. 

For a more comprehensive treatment of the problem, including vertex arrivals, integral solutions, and disposal models, we refer the reader to~\cite{BBDFGK26}. 

\paragraph{Our result.} 

We give a deterministic primal-dual algorithm, \textsc{Host-or-Halve} (\HoH), that maintains simultaneously a solution $\vy^t$ to online fractional vertex cover and a solution $\vm^t$ to online fractional matching. On forests, the former is $(4 - \sqrt{5})$-competitive, where $4 - \sqrt{5} \approx 1.764$, and the latter is $1/(4 - \sqrt{5})$-competitive, where $1/(4 - \sqrt{5}) \approx 0.566$ (\cref{thm:main}). 

This improves both over the ratio of $1.833$ for online fractional vertex cover~\cite{BBDFGK26} and over the ratio of $0.555$ for online fractional matching~\cite{BuSeTk19}. 

\paragraph{Independent work.}

Independently of this paper, Bosek, Gutowski, Putra and Zych-Pawlewicz obtain better bounds by constructing a primal-dual algorithm for forests that is simultaneously $7/4$-competitive for online fractional vertex cover and $4/7$-competitive for online fractional matching~\cite{BoGuPZ26}. 

\paragraph{AI disclosure.} 

The function~$\gs$ of \cref{sec:golden} was optimized with the help of Claude Opus 5.5. The resulting function and its properties were checked manually by the author, who takes full responsibility for the whole content of this paper.


\section{Algorithm Host-or-Halve}
\label{sec:alg}

An arriving edge with exactly one endpoint of degree~$0$ (degrees are taken before the edge arrives) is a \emph{leaf attachment}. Then, the vertex with degree~$0$ is the \emph{leaf} and the other vertex is the \emph{host}. Every other arriving edge is a \emph{merge}. A merge is \emph{trivial} if both its endpoints have degree~$0$, and \emph{non-trivial} otherwise; in a~forest, a non-trivial merge joins two trees of the revealed forest.

\paragraph{Intuition.}

Before we formally define the algorithm, we give main ideas behind its construction and analysis. When algorithm is performing a merge, it raises the weights of both endpoints at the same pace. On the other hand, when it is performing a leaf attachment, it raises the weight of the host vertex from $w$ to $g(w)$, where $g$ is a carefully chosen non-decreasing function satisfying $g(w) \geq w$. The weight of the attached leaf is set to $1 - g(w)$, so that the edge is covered. The competitive ratio is then expressed in terms of the function~$g$. The intuition behind favoring the host is that more leaves may be later attached to it, and each such attachment is cheaper when the weight of the host is larger.

For the analysis, a rough observation is that the local performance of the algorithm is best on trivial merges, but worst on non-trivial merges. We note however that in a forest the number of non-trivial merges is at most that of trivial merges. Hence, in our analysis, we assign ``credit'' to trivial merges, which can be used to compensate for the cost of non-trivial merges.

\paragraph{Notation.}

For $t \geq 0$, let $E_t$ be the set of edges revealed in steps $1, \dots, t$, and let $V_t$ be the set of vertices incident to at least one edge of~$E_t$. For $t \geq 1$, let $e_t$ be the edge that arrives in step~$t$, and let $\Delta_t \size{\vy} \define \size{\vy^t} - \size{\vy^{t-1}}$ be the weight added to the vertex cover in step~$t$. Let $\kappa_t$ be the number of components of the graph revealed in steps $1, \dots, t$; note that $\kappa_0 = 0$. Let $\Delta_t \kappa \define \kappa_t - \kappa_{t-1}$. Note that $\Delta_t \kappa$ is $1$ for a trivial merge, $0$ for a leaf attachment, and is $-1$ for a non-trivial merge. Note that this value of $\Delta_t \kappa$ for a non-trivial merge requires that the graph is a forest and, in fact, it is the only place where the forest assumption is used.

\begin{definition}
    \label{def:alg}
    Let $g \colon [0,1] \to [0,1]$ be a non-decreasing function such that $g(w) \geq w$ for every~$w$. The algorithm \textsc{Host-or-Halve}(g) ($\HoH(g)$) maintains a vector~$\vy$ of vertex weights and a vector~$\va$ of edge weights, both initially~$\zero$, and it outputs the cover $\vy^t$ and the matching $\vm^t \define \va^t/\rho_g$, where $\rho_g$ is the constant defined below. Upon arrival of an edge $e_t = (u,v)$, $\HoH(g)$ first updates the weights of $u$ and~$v$ in the following way.
    \begin{description}
        \item[leaf attachment.] If exactly one endpoint, say $v$, has degree~$0$, we call $u$ the \emph{host}. Let $w$ be the current weight of~$u$. $\HoH(g)$ sets $y_u \define g(w)$ and $y_v \define 1 - g(w)$.
        \item[merge.] Otherwise, let $\deff \define 1 - y_u - y_v$ be the \emph{deficiency} of~$(u,v)$. If $\deff > 0$, the weights of both endpoints increase by $\deff/2$. If $\deff \leq 0$, no weight changes.
    \end{description}
    Then, $\HoH(g)$ sets
    \[
        a_{e_t} \define \max \set{0, \; \Delta_t \size{\vy} + \beta_g \cdot \Delta_t \kappa} ,
    \]
    where $\beta_g \geq 0$ is the constant defined below, and it never modifies $a_{e_t}$ later.
\end{definition}

\begin{figure}[t]
    \centering
    \begin{tikzpicture}[x=1cm,y=1cm,font=\small,
        v/.style={circle,draw,fill=white,inner sep=2pt},
        nv/.style={circle,draw=figA,fill=figA!25,inner sep=2pt},
        e/.style={thick},
        ne/.style={very thick,figB},
        comp/.style={draw=gray,rounded corners=6pt,densely dotted},
        ]
        \begin{scope}
            \node[nv] (s) at (0,0) {};
            \node[nv] (t) at (1.5,0) {};
            \draw[ne] (s) -- (t);
            \node[below=4pt] at (s) {$0 \to \tfrac12$};
            \node[below=4pt] at (t) {$0 \to \tfrac12$};
            \node at (0.75,-1.6) {(a) trivial merge};
        \end{scope}
        \begin{scope}[xshift=4.2cm]
            \node[v] (b) at (-1.0,0.9) {};
            \node[v] (c) at (-1.0,-0.9) {};
            \node[nv] (s) at (0,0) {};
            \node[nv] (t) at (1.5,0) {};
            \draw[e] (s) -- (b); \draw[e] (s) -- (c);
            \draw[ne] (s) -- (t);
            \node[above right=3pt and -2pt] at (s) {$w \to g(w)$};
            \node[below=4pt] at (t) {$0 \to 1 - g(w)$};
            \node at (0.25,-1.6) {(b) leaf attachment};
        \end{scope}
        \begin{scope}[xshift=9.2cm]
            \node[v] (a1) at (-1.0,0.9) {}; \node[v] (a2) at (-1.0,-0.9) {};
            \node[nv] (u) at (0,0) {};
            \draw[e] (u) -- (a1); \draw[e] (u) -- (a2);
            \node[v] (b1) at (4.0,0.9) {}; \node[v] (b2) at (4.0,-0.9) {};
            \node[nv] (w) at (3.0,0) {};
            \draw[e] (w) -- (b1); \draw[e] (w) -- (b2);
            \draw[ne] (u) -- (w);
            \node[below right=3pt and -2pt] at (u) {$\tfrac15 \to \tfrac25$};
            \node[below left=3pt and -2pt] at (w) {$\tfrac25 \to \tfrac35$};
            \draw[comp] (-1.35,-1.15) rectangle (0.45,1.15);
            \draw[comp] (2.55,-1.15) rectangle (4.35,1.15);
            \node at (1.5,-1.6) {(c) non-trivial merge, deficiency $\deff = 2/5$};
        \end{scope}
    \end{tikzpicture}
    \caption{The three kinds of arriving edges (thick) in the analysis of the algorithm $\HoH(g)$ of \cref{def:alg}, with the weights before and after the arrival.}
    \label{fig:alg}
\end{figure}
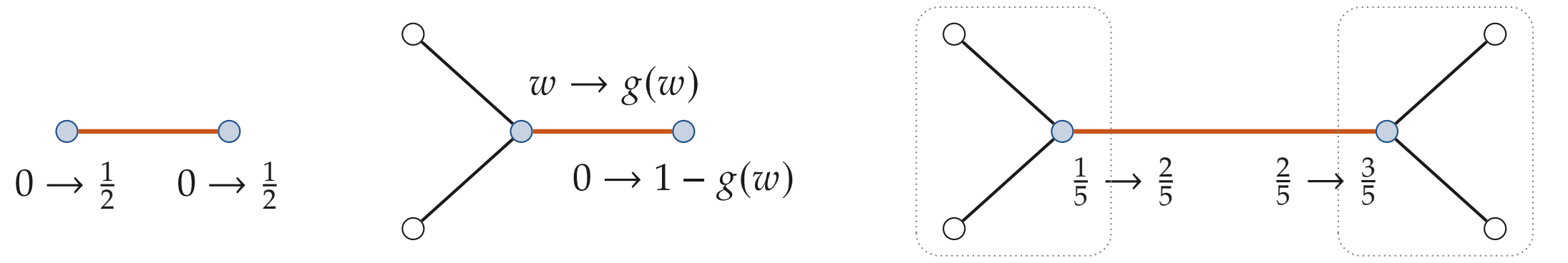

Clearly, $\vy^t$ is a feasible solution to the vertex cover problem; what remains to be shown is that $\vm^t$ is a feasible solution to the matching problem. We then observe that the ``credit'' terms $\beta_g \cdot \Delta_t \kappa$ sum up to a non-negative value, which leads to $\size{\vy^t} \leq \rho_g \cdot \size{\vm^t}$. This will show the competitive ratio of $\rho_g$ for both the vertex cover and the matching problems. 

To proceed with the formal analysis, we first need to define $\beta_g$ and $\rho_g$.

\paragraph{Parameters.}

We define
\[
    T_g(w) \define \textstyle\sum_{i \geq 0} \bigl( 1 - g^i(w) \bigr)
\]
where $g^i$ is the $i$-th iterate of~$g$ and $g^0(w) = w$. By the definition, $T_g(w) = (1 - w) + T_g(g(w))$, and $T_g(1) = 0$ as $g(1) = 1$. Thus, $T_g(w)$ is the total weight that $\HoH(g)$ adds to the cover while new leaves are attached, one after another, to a vertex of initial weight~$w$. 

In the analysis, we restrict attention to functions~$g$ for which $T_g(w)$ is finite for every~$w$. For such functions, let
\begin{align*}
    \beta_g & \define \sup \left\{ \bigl( T_g(w') + 2w' \bigr) - \bigl( T_g(w) + 2w \bigr) 
        \suchthat 0 \leq w \leq w' \leq (1 + w)/2 \right\}, \\
    \gamma_g & \define \sup \left\{ 1 - w + T_g(1 - g(w)) \suchthat 0 \leq w \leq 1 \right\} , \\
    \rho_g & \define \max \set{\gamma_g, \; 1 + \beta_g + T_g(1/2)} .
\end{align*}
Taking $w' = w$ shows that $\beta_g \geq 0$. The choice and meaning of the values of $\beta_g$, $\gamma_g$, and $\rho_g$ will become clear in the analysis of $\HoH(g)$ in \cref{sec:dual,sec:feasible}.

\subsection{Competitive ratios}
\label{sec:dual}

The main technical part of the analysis is to show that the matching $\vm^t = \va^t/\rho_g$ is feasible (cf.~\cref{lem:feasible}). The proof is postponed to \cref{sec:feasible}. 

\begin{lemma}[dual feasibility]
    \label{lem:feasible}
    Fix a function~$g$ as in \cref{def:alg} with finite $T_g$. For every step $t \geq 0$, the vector $\va^t/\rho_g$ is a feasible fractional matching of~$E_t$.
\end{lemma}

\begin{theorem}
    \label{thm:main}
    Fix a function~$g$ as in \cref{def:alg} with finite $T_g$, and assume that the input graph is a forest. Then, the cover $\vy^t$ maintained by $\HoH(g)$ is $\rho_g$-competitive for online fractional vertex cover, and the matching $\vm^t$ maintained by $\HoH(g)$ is $(1/\rho_g)$-competitive for online fractional matching.
\end{theorem}

\begin{proof}
    Fix a step $t \geq 0$. The cover $\vy^t$ is feasible by the definition of $\HoH(g)$, and the matching $\vm^t$ is feasible by \cref{lem:feasible}. Thus,
    \begin{align*}
        \Opt_t
        & \leq \size{\vy^t}
            && \text{(as $\vy^t$ is feasible)} \\
        & \leq \size{\vy^t} + \beta_g \cdot \kappa_t
            && \text{(as $\beta_g \geq 0$ and $\kappa_t \geq 0$)} \\
        & = \textstyle\sum_{s=1}^{t} \bigl( \Delta_s \size{\vy} + \beta_g \cdot \Delta_s \kappa \bigr)
            && \text{(as $\kappa_0 = 0$)} \\
        & \leq \textstyle\sum_{s=1}^{t} a_{e_s}
            && \text{(by the definition of $a_{e_s}$)} \\
        & = \rho_g \cdot \size{\vm^t}
            && \text{(as $\vm^t = \va^t / \rho_g$)} \\
        & \leq \rho_g \cdot \Opt_t .
            && \text{(as $\vm^t$ is feasible)}
    \end{align*}
    Hence, $\size{\vy^t} \leq \rho_g \cdot \Opt_t$ and $\size{\vm^t} \geq \Opt_t / \rho_g$, which concludes the proof.
\end{proof}


\subsection{Dual feasibility}
\label{sec:feasible}

In this section, we prove \cref{lem:feasible}. We fix a~function~$g$ as in \cref{def:alg} with finite $T_g$, and we assume that the input graph is a forest. We write $T = T_g$, $\rho = \rho_g$, $\beta = \beta_g$ and $\gamma = \gamma_g$. As $g$ is non-decreasing, so is every iterate~$g^i$, and thus $T$ is non-increasing. 

For $t \geq 0$, let 
\[ 
    \textstyle L_t(x)
    \define \sum\nolimits_{e \in E_t \suchthat x \in e} a^t_e 
    = \sum\nolimits_{s \leq t \suchthat x \in e_s} a_{e_s}
\]
be the \emph{load} of a vertex~$x$ at time~$t$. As $\va^t \geq \zero$, to show that $\va^t/\rho$ is a feasible fractional matching of~$E_t$, it suffices to show that $L_t(x) \leq \rho$ for every vertex~$x$.

\paragraph{Free part of a dual constraint.} For $t \geq 0$ and a vertex~$x$, let $y^t_x$ be the weight of~$x$ after step~$t$. For $x \in V_t$, we show that $T(y^t_x)$ lower-bounds the slack of the inequality $L_t(x) \leq \rho$, i.e., we show that $L_t(x) + T(y^t_x) \leq \rho$. In other words, $T(y^t_x)$ is the ``budget'' that vertex~$x$ keeps for the edges that may arrive at~$x$ later. The next two lemmas bound the dual value of an edge at each of its endpoints: \cref{lem:first} for an endpoint that the edge reveals, and \cref{lem:later} for an endpoint that already has positive degree.

\begin{lemma}
    \label{lem:first}
    For every step $t \geq 1$ and each degree-$0$ endpoint~$x$ of~$e_t$, it holds that $a_{e_t} \leq \rho - T(y^t_x)$.
\end{lemma}

\begin{proof}
    Let $(x,u) = e_t$. As $x$ has degree~$0$, edge $e_t$~is not a non-trivial merge, so $\Delta_t \kappa \geq 0$. As also $\Delta_t \size{\vy} \geq 0$, we have $a_{e_t} = \Delta_t \size{\vy} + \beta \cdot \Delta_t \kappa$, and it remains to show that $\Delta_t \size{\vy} + \beta \cdot \Delta_t \kappa \leq \rho - T(y^t_x)$. We consider two cases depending on the degree of~$u$.
    \begin{itemize}
        \item $u$ has degree~$0$. Then $e_t$ is a trivial merge, $y^t_x = 1/2$ and $\Delta_t \kappa = 1$, so
            \begin{align*}
                \Delta_t \size{\vy} + \beta \cdot \Delta_t \kappa
                & = 1/2 + 1/2 + \beta \\
                & \leq \rho - T(1/2).
                    && \text{(by the definition of $\rho$)}
            \end{align*}
        \item $u$ has positive degree. Then $e_t$ is a leaf attachment, $x$ is the leaf, and $u$ is the host. Let $w \define y^{t-1}_u$. Then $y^t_u = g(w)$ and $y^t_x = 1 - g(w)$, so
            \begin{align*}
                \Delta_t \size{\vy} + \beta \cdot \Delta_t \kappa
                & = (g(w) - w) + (1 - g(w)) + 0 \\
                & = 1 - w \\
                & \leq \gamma - T(1 - g(w))
                    && \text{(by the definition of $\gamma$)} \\
                & \leq \rho - T(y^t_x).
                    && \text{(by the definition of $\rho$)} 
                    \qedhere
            \end{align*}
    \end{itemize}
\end{proof}

\begin{lemma}
    \label{lem:later}
    For every step $t \geq 1$ and each positive-degree endpoint~$x$ of~$e_t$, it holds that $a_{e_t} \leq T(y^{t-1}_x) - T(y^t_x)$.
\end{lemma}

\begin{proof}
    Let $(x,u) = e_t$. As $y^t_x \geq y^{t-1}_x$ and $T$ is non-increasing, $0 \leq T(y^{t-1}_x) - T(y^t_x)$. As $a_{e_t} = \max \set{0, \; \Delta_t \size{\vy} + \beta \cdot \Delta_t \kappa}$, it remains to show that $\Delta_t \size{\vy} + \beta \cdot \Delta_t \kappa \leq T(y^{t-1}_x) - T(y^t_x)$. We consider two cases depending on the degree of~$u$.
    \begin{itemize}
        \item $u$ has degree~$0$. Then $e_t$ is a leaf attachment, $x$ is the host, and $u$ is the leaf. Let $w \define y^{t-1}_x$. Then $y^t_x = g(w)$, $y^t_u = 1 - g(w)$ and $\Delta_t \kappa = 0$, so
            \begin{align*}
                \Delta_t \size{\vy} + \beta \cdot \Delta_t \kappa
                & = (g(w) - w) + (1 - g(w)) + 0 \\
                & = 1 - w \\
                & = T(w) - T(g(w))
                    && \text{(by the definition of $T$)} \\
                & = T(y^{t-1}_x) - T(y^t_x) .
            \end{align*}
        \item $u$ has positive degree. Then $e_t$ is a non-trivial merge, and $\Delta_t \kappa = -1$. Let $w \define y^{t-1}_x$ and $\delta \define \max \set{0,\, 1 - y^{t-1}_x - y^{t-1}_u}/2$. The weights of both endpoints increase by~$\delta$, so $\Delta_t \size{\vy} = 2 \cdot \delta$. As $\delta \leq (1-w)/2$, we have $w \leq w + \delta \leq (1 + w)/2$. Thus, the definition of~$\beta$ applied with $w' = w + \delta$ yields $\beta \geq T(w + \delta) + 2 \cdot (w + \delta) - T(w) - 2 \cdot w$. Hence,
            \begin{align*}
                \Delta_t \size{\vy} + \beta \cdot \Delta_t \kappa
                & = 2 \cdot \delta - \beta \\
                & \leq T(w) - T(w + \delta)
                    && \text{(by the bound on $\beta$)} \\
                & = T(y^{t-1}_x) - T(y^t_x) .
                    && \qedhere
            \end{align*}
    \end{itemize}
\end{proof}

\begin{proof}[Proof of \cref{lem:feasible}]
    Fix a step $t \geq 0$ and a vertex~$x$. If $x \notin V_t$, then $L_t(x) = 0 \leq \rho$. Otherwise, let $s \leq t$ be the first step with $x \in e_s$. Then,
    \begin{align*}
        L_t(x)
        & = a_{e_s} + \textstyle\sum_{s < q \leq t \suchthat x \in e_q} a_{e_q} \\
        & \leq \rho - T(y^s_x) + \textstyle\sum_{s < q \leq t \suchthat x \in e_q} \bigl( T(y^{q-1}_x) - T(y^q_x) \bigr)
            && \text{(by \cref{lem:first,lem:later})} \\
        & = \rho - T(y^t_x)
            && \text{(as $y_x$ changes only in steps $q$ with $x \in e_q$)} \\
        & \leq \rho.
            && \text{(as $T \geq 0$)}  \qedhere
    \end{align*}
\end{proof}


\subsection{Optimizing constants}
\label{sec:golden}

Let $\varphi = (1 + \sqrt{5})/2$ be the golden ratio. Let $c \define 1/\varphi = (\sqrt{5} - 1)/2 \approx 0.618$. Note that $(1 + c) \cdot c = 1$. In this section, we show that the function 
\[ 
    \gs(w) = \begin{cases} 
        c \cdot (1 + w) & \text{if }w \leq c, \\
        1 & \text{if } w > c \\ 
    \end{cases}
\] 
gives $\rho_{\gs} \leq 3 - 2c = 4 - \sqrt{5}$. 

It is straightforward to verify that $\gs$ satisfies the conditions of \cref{def:alg}. Note that $\gs(w) \geq c$ for every~$w$, so $\gs(\gs(w)) = 1$ for every~$w$, and hence
\[
    T_{\gs}(w) = (1 - w) + (1 - \gs(w)) = 2 - w - \gs(w) \qquad \text{for every } w \in [0,1] ,
\]
that is,
\[
    T_{\gs}(w) = \begin{cases}
        2 - c - (1 + c) \cdot w & \text{if } w \leq c , \\
        1 - w & \text{if } w > c .
    \end{cases}
\]

\begin{lemma}
    \label{lem:golden_beta}
    It holds that $\beta_{\gs} \leq (1-c)/2$.
\end{lemma}

\begin{proof}
    We have $T_{\gs}(w) + 2w = 2 + w - \gs(w)$, and thus
    \[
        \beta_{\gs} = \sup \set{ w' - \gs(w') - w + \gs(w) \suchthat 0 \leq w \leq w' \leq (1 + w)/2 } .
    \]
    The function $b \mapsto b - \gs(b)$ is non-decreasing, as $\gs$ is continuous with slope at most $c < 1$. Hence, for a fixed~$w$, the expression in the supremum is non-decreasing in~$w'$. Thus, the supremum is attained at $w' = (1 + w)/2$, and $\beta_{\gs} = \sup_{w \in [0,1]} \phi(w)$, where
    \[
        \phi(w) \define (1 - w)/2 + \gs(w) - \gs \bigl( (1 + w)/2 \bigr) .
    \]
    The function~$\gs$ is concave, as its slope drops from~$c$ to~$0$. On $[0,c]$, $\gs(w)$ is affine and $-\gs((1 + w)/2)$ is convex, so $\phi$ is convex and is maximized at an endpoint. On $[c,1]$, both $\gs(w)$ and $\gs((1 + w)/2)$ equal~$1$, so $\phi(w) = (1 - w)/2$ is decreasing. Hence, $\phi$ is maximized at $w = 0$ or at $w = c$. As $1/2 < c < (1 + c)/2$, we have $\phi(0) = 1/2 + c - 3c/2 = (1 - c)/2$ and $\phi(c) = (1 - c)/2 + 1 - 1 = (1 - c)/2$. Thus, $\beta_{\gs} \leq (1-c)/2$.
\end{proof}

\begin{lemma}
    \label{lem:golden}
    It holds that $\rho_{\gs} \leq 3 - 2c = 4 - \sqrt{5}$.
\end{lemma}

\begin{proof}
    By the definition of $\rho_{\gs}$, it suffices to show that $\gamma_{\gs} \leq 3 - 2c$ and $1 + \beta_{\gs} + T_{\gs}(1/2) \leq 3 - 2c$. For the former bound, fix any $w \in [0,1]$, and observe that
    \begin{align*}
        (1 - w) + T_{\gs}(1 - \gs(w)) 
        & = (1 - w) + 2 - c - (1 + c) \cdot (1 - \gs(w)) 
            && \text{(as $1 - \gs(w) \leq 1 - c < c$)} \\
        & = 2 - 2c - w + (1 + c) \cdot \gs(w) \\
        & \leq 2 - 2c - w + (1 + w) 
            && \text{(as $(1 + c) \cdot \gs(w) \leq 1 + w$)} \\
        & = 3 - 2c .
    \end{align*}
    Hence, $\gamma_{\gs} \leq 3 - 2c$. For the latter bound, we have 
    \begin{align*}
        1 + \beta_{\gs} + T_{\gs}(1/2) 
        & \leq 1 + (1-c)/2 + 2 - c - (1 + c)/2
            && \text{(by \cref{lem:golden_beta} and as $1/2 \leq c$)} \\
        & = 3 - 2c. 
            && \qedhere
    \end{align*}
\end{proof}

Applying function $\gs$ to \cref{thm:main} and using $\rho_{\gs} \leq 4 - \sqrt{5}$ by \cref{lem:golden}, immediately gives the main result of this paper.

\begin{corollary}
    \label{cor:golden}
    For the function~$\gs$ defined above, the algorithm $\HoH(\gs)$ executed on an input graph that is a forest produces a pair of feasible solutions $\vy^t$ and $\vm^t$. The fractional vertex cover $\vy^t$ has competitive ratio at most $4 - \sqrt{5} \approx 1.764$ and the matching $\vm^t$ has competitive ratio at least $1/(4 - \sqrt{5}) \approx 0.566$.
\end{corollary}

\bibliography{references}

\newcommand{\etalchar}[1]{$^{#1}$}
\begin{thebibliography}{GKM{\etalchar{+}}19}

\bibitem[BBD{\etalchar{+}}26]{BBDFGK26}
J{\'{u}}lia Balig{\'{a}}cs, Bart{\l}omiej Bosek, Yann Disser, Andreas~Emil
  Feldmann, Grzegorz Gutowski, Katarzyna K{\k{e}}pi{\'{n}}ska, Pawe{\l} Putra,
  and Anna Zych{-}Pawlewicz.
\newblock Online and incremental fractional vertex cover on trees.
\newblock In {\em Proc. 34th European Symp. on Algorithms (ESA)}, volume 388 of
  {\em LIPIcs}, pages 158:1--158:15. Schloss Dagstuhl, 2026.
\newblock \href {https://doi.org/10.4230/LIPICS.ESA.2026.158}
  {\path{doi:10.4230/LIPICS.ESA.2026.158}}.

\bibitem[BGPZ26]{BoGuPZ26}
Bart{\l}omiej Bosek, Grzegorz Gutowski, Pawe{\l} Putra, and Anna
  Zych{-}Pawlewicz.
\newblock Tight competitive bounds for online matching on forests.
\newblock Unpublished manuscript, 2026.

\bibitem[BST19]{BuSeTk19}
Niv Buchbinder, Danny Segev, and Yevgeny Tkach.
\newblock Online algorithms for maximum cardinality matching with edge
  arrivals.
\newblock {\em Algorithmica}, 81(5):1781--1799, 2019.
\newblock \href {https://doi.org/10.1007/S00453-018-0505-7}
  {\path{doi:10.1007/S00453-018-0505-7}}.

\bibitem[GKM{\etalchar{+}}19]{GaKMSW19}
Buddhima Gamlath, Michael Kapralov, Andreas Maggiori, Ola Svensson, and David
  Wajc.
\newblock Online matching with general arrivals.
\newblock In {\em Proc. 60th IEEE Symp. on Foundations of Computer Science
  (FOCS)}, pages 26--37, 2019.
\newblock \href {https://doi.org/10.1109/FOCS.2019.00011}
  {\path{doi:10.1109/FOCS.2019.00011}}.

\bibitem[TZ26]{TanZha26}
Zhihao~Gavin Tang and Yuhao Zhang.
\newblock A tight bound on online vertex cover under edge arrivals.
\newblock arXiv:2608.04994, 2026.
\newblock \href {https://arxiv.org/abs/2608.04994} {\path{arXiv:2608.04994}}.

\end{thebibliography}

\end{document}